\documentclass[preprint,showpacs,preprintnumbers,amsmath,amssymb]{revtex4}

\usepackage{graphicx}
\usepackage{dcolumn}
\usepackage{bm}

\begin{document}


\title{Quantum Arrival Time Formula from Decoherent Histories}

\author{J.J.Halliwell}%
\email{j.halliwell@ic.ac.uk}

\author{J.M.Yearsley}
\email{james.yearsley03@imperial.ac.uk}

\affiliation{Blackett Laboratory \\ Imperial College \\ London SW7
2BZ \\ UK }



\begin{abstract}
In the arrival time problem in quantum mechanics, a standard
formula that frequently emerges as the probability for crossing
the origin during a given time interval is the current integrated
over that time interval. This is semiclassically correct but can
be negative due to backflow. Here, we show that this formula
naturally arises in a decoherent histories analysis of the arrival
time problem.
For a variety of initial states, we show that histories crossing
during different time intervals are approximately decoherent. Probabilities may therefore
be assigned and coincide with the standard formula (in a semiclassical
approximation), which
is therefore positive for these states.
However, for initial states for which there is
backflow, we show that there cannot be decoherence of
histories, so probabilities may not be assigned.

\end{abstract}



\maketitle

\newcommand\beq{\begin{equation}}
\newcommand\eeq{\end{equation}}
\newcommand\bea{\begin{eqnarray}}
\newcommand\eea{\end{eqnarray}}

\def\A{{\cal A}}
\def\D{\Delta}
\def\H{{\cal H}}
\def\E{{\cal E}}
\def\p{\partial}
\def\la{\langle}
\def\ra{\rangle}
\def\ria{\rightarrow}
\def\x{{\bf x}}
\def\y{{\bf y}}
\def\k{{\bf k}}
\def\q{{\bf q}}
\def\p{{\bf p}}
\def\P{{\bf P}}
\def\r{{\bf r}}
\def\s{{\sigma}}
\def\a{\alpha}
\def\b{\beta}
\def\e{\epsilon}
\def\U{\Upsilon}
\def\G{\Gamma}
\def\om{{\omega}}
\def\Tr{{\rm Tr}}
\def\ih{{ \frac {i} { \hbar} }}
\def\trho{{\rho}}

\def\au{{\underline \alpha}}
\def\bu{{\underline \beta}}
\def\pp{{\prime\prime}}
\def\id{{1 \!\! 1 }}
\def\half{\frac {1} {2}}

\def\jjh{j.halliwell@ic.ac.uk}


In the quantum arrival time problem, one considers a free particle
in an initial state $ | \psi \rangle $
localized in $x>0$ consisting entirely of negative momenta, and asks
for the probability $p(t_1,t_2)$ that the particle crosses the origin
during the time interval $[t_1,t_2]$. There are many approaches
to this specific problem and to time in quantum theory generally \cite{time,Kij,All}.
But a frequently discussed candidate for this probability is the integrated current,
\bea
p (t_1, t_2) &=& \int_{t_1}^{t_2} dt \ J(t)
\nonumber \\
&=& \langle \psi | C | \psi \rangle
\label{1}
\eea
where
\bea
C &=& \int_{t_1}^{t_2} dt  \ \frac { (-1) } {2 m } \left( \hat p \delta ( \hat x ) + \delta (
\hat x ) \hat p \right)
\nonumber \\
&=& \theta (\hat x (t_1) ) - \theta ( \hat x (t_2) )
\label{2}
\eea
(For convenience we work in units in which $ \hbar =
1$). This is sensible classically and correctly normalized
as $t_2 \rightarrow \infty $ with $t_1 = 0$ (assuming that
all of the state ends up in $x<0$ at large times). But it can be negative in the
quantum case for certain states consisting of superpositions of
different momenta. This genuinely quantum phenomenon is called backflow
and arises because the operator $C$, positive classically, has negative
eigenvalues  \cite{cur,BrMe,back}.

The flux Eq.(\ref{1}), and simple variants of it, are measurable \cite{flux}.
An interesting problem is therefore to derive Eq.(\ref{1}), at least in some
approximation, from an underlying axiomatic scheme
or specific model of quantum measurements. Here, we present such a derivation using
the decoherent histories approach to quantum theory, which is naturally adapted to
questions of this type \cite{GeH1,Gri,Omn,Hal2,DoH,Ish}. This derivation establishes the conditions under which one
would expect the formula Eq.(\ref{1}) to hold. In particular, we shall see that
it is not expected to hold in precisely those situations when there is backflow.

A number of previous authors have used the decoherent histories approach to analyse
this and similar problems involving time in a non-trivial way
\cite{Har,MiH,HaZa,YaT,AnSa}, but none of them make contact with the standard result
Eq.(\ref{1}).

We begin by making some simple observations that capture the essence of what
we do in the rest of the paper. The classical analogue of the
hermitian operator $C$ defined above is a phase space function
which is $1$ or $0 $ (for initial states of the form considered
here). Classically, there is therefore no difference between $C$ and $C^2$.
Therefore, in the quantum theory, one could just as easily propose
$ \langle C^2 \rangle $ instead of $ \langle C \rangle $ as the crossing probability, since they both have the same classical limit. Their difference
may be written
\beq
\langle \psi | C^2 | \psi \rangle  - \langle \psi | C| \psi  \rangle = - \langle \psi
|C ( 1 - C)| \psi \rangle
\label{3}
\eeq
The right-hand side is an overlap between the state $ C | \psi \rangle $, representing
crossing during the given time and interval, and $ (1 - C) | \psi \rangle$, representing
not crossing during that time interval (so crossing at another time). It therefore
represents the interference between crossing and not crossing. When there is no
interference we have
\beq
\langle C \rangle = \langle C^2 \rangle \ge  0
\label{4}
\eeq
and the sometimes negative number $ \langle C \rangle $ is then assured to be positive.
On the other hand, when there is backflow, $\langle C \rangle < 0 $, which implies that
the right-hand
side of Eq.(3) must be non-zero. This shows that backflow is strongly linked to
interference effects. These heuristic comments have a natural setting in the decoherent
histories approach to quantum theory, which we now describe.

Alternatives at fixed moments of time in quantum theory are
represented by a set of projection operators $\{ P_a \}$,
satisfying the conditions
\bea
\sum_a P_a &=& 1
\\
P_a P_b &=& \delta_{ab} P_a \eea where we take $a$ to run over
some finite range. In the decoherent histories approach to quantum
theory \cite{GeH1,Gri,Omn,Hal2,DoH,Ish}, the simplest type of history, a homogenous history, is
represented by a class operator $C_{\a}$ which is a time-ordered
string of projections
\beq
C_{\a} = P_{a_n} (t_n) \cdots P_{a_1}
(t_1)
\label{1.3}
\eeq
Here the projections are in the Heisenberg
picture and $ \a $ denotes the string $ (a_1, \cdots a_n)$.
The theory also allows so-called inhomogenous histories, whose
class operators are sums of the
the basic class operator Eq.(\ref{1.3}).
All class operators satisfies the condition
\beq
\sum_{\a} C_{\a} = 1
\label{1.4}
\eeq
Probabilities are assigned to histories via the formula
\beq
p(\a) = {\rm Tr} \left( C_{\a} \rho C_{\a}^{\dag} \right)
\label{1.6}
\eeq
Probabilities assigned in this way do not necessarily obey the
probability sum rules, because of quantum interference.
Restrictions on the permissable sets of histories are therefore necessary
to ensure that there are no interference effects.
To this end, we introduce the decoherence functional
\beq
D(\a, \b) = {\rm Tr} \left( C_{\a} \rho C_{\b}^{\dag} \right)
\eeq
which may be thought of as a measure of interference between pairs of histories
and require that sets of histories satisfy the condition of
decoherence, which is
\beq
D(\a, \b) = 0, \ \ \ \a \ne \b
\eeq
This ensures that all probability sum rules are satisfied.

It is useful to define the quasi-probability
\beq
q(\a) = {\rm Tr} \left( C_{\a} \rho \right)
\label{1.17}
\eeq
Because it is linear in the $C_{\a}$, this quantity satisfies the
probability sum rules and sums to $1$, but it is not in general a real number.
However, it is closely related to the probabilities Eq.(\ref{1.6}), because
Eq.(\ref{1.4}) implies that
\beq
q (\a ) = p(\a ) + \sum_{\b, \b \ne \a}
D (\a, \b)
\label{1.18}
\eeq
This means that when there is decoherence the probabilities
are given by the simpler expression
\beq
p(\a ) = q (\a)
\label{1.20}
\eeq
Decoherence therefore ensures that $q(\a)$ is real and positive, even though it
is not in general. (Note also that requiring $q(\a)$ to be real and positive is not enough
to guarantee decoherence, although if $q(\a)$ is not real, or is real but negative,
then there cannot be decoherence).

These features of the decoherent histories together with the heuristic argument
given above strongly suggest that the standard formula for arrival time probability,
Eq.(\ref{1}) is in fact a {\it quasi-probability} of the form Eq.(\ref{1.17}),
with class operators given by expressions of the form Eq.(\ref{2}). This explains
why Eq.(\ref{1}) gives reasonable answers in some circumstances but not in
others. Decoherence of histories is the missing element required
to understand the regime of validity of the formula, and the negativity of Eq.(\ref{1})
when there is backflow is seen to be
a consequence of lack of decoherence.
To substantiate these claims, we need to explicitly construct
the class operators, confirm that they are of the form Eq.({\ref{2}) and
also confirm that certain simple states of interest satisfy the decoherence condition.

We first construct the class operators.
We consider an initial state at $t=0$ and suppose the state crosses the origin
during a large time interval $[0, \tau]$. It is useful to introduce a discrete set of times
$t_k = k \epsilon $, where $ k = 0,1 \cdots n $ and $ \tau = n \epsilon $.
We also define the projection operators
\bea
P &=& \theta ( \hat x )
\\
\bar P &=& 1 - P = \theta ( - \hat x )
\eea

Consider first the class operator for remaining in $x>0$ (i.e., not crossing $x=0$) during
the time interval $ [ 0, \tau ]$.
We assert that the appropriate class operator is
\beq
C_{nc} = P (t_n) \cdots P(t_2) P(t_1)
\label{17}
\eeq
This corresponds to the statement that the particle is in $x>0$ at the discrete set of
times $t_1, t_2, \cdots t_n$,
but its location is unspecified at intermediate times. It might seem that it is appropriate to take
the limit $ \epsilon \rightarrow 0$, thereby obtaining a class operator ensuring that the
particle is in $x>0$ at {\it every} time in the interval $ [0, \tau]$. However, the
resulting object has the form
\beq
C_{nc} \rightarrow e^{ i H \tau } g_r ( \tau, 0 )
\eeq
where $g_r $ is the restricted propator in $x>0$.  It actually describes
unitary propagation in $x>0$, as may be seen from the representation
\beq
g_r (\tau, 0 ) = P \exp \left( - i P H P \tau \right)
\eeq
This means that the state never in fact leaves $x>0$ (and in fact $g_r$ describes
the situation in which the incoming state undergoes total reflection at the origin)
\cite{Wall,Sch2}.
This is clearly unphysical for the arrival time problem and is essentially
the Zeno effect:  monitoring the system too closely prevents it from making
any physical interesting transitions \cite{Zeno}. To avoid the Zeno effect, we must
leave $ \epsilon $ finite in Eq.(\ref{17}). Studies of the Zeno effect suggest that
the important timescale is the Zeno time
\beq
t_z = \frac {1} { \Delta H}
\label{18}
\eeq
and that significant reflection is avoided if $ \epsilon > t_z$.

Consider now the class operator for crossing during a time interval. The notion of
crossing a surface of constant $x$ in quantum mechanics is a subtle one. In a path
integral construction, for example, the notion of crossing a surface of constant
$t$ is well-defined since the paths cross such a surface once and only once.
The notion of crossing a surface of constant $x$, however, is not well-defined --
a path from one side of the surface to the other will typically cross the surface
an infinite number of times \cite{Har}. However, notions of crossing
that are well-defined in path integral constructions are the {\it first} crossing
and {\it last} crossing and we will use this to guide us here \cite{PDX}.


We construct the first crossing class operator by partitioning the
histories according to whether they are in $x<0$ or $ x>0$ at the
discrete set of times $t_k$ and noting that the class operators
must sum to the identity. We write
\bea
1 &=& \bar P(t_1) + P(t_1)
\nonumber \\
&=& \bar P(t_1) + \bar P (t_2) P(t_1) + P (t_2) P(t_1)
\eea
Repeating inductively, we obtain
\bea
1 &=& \bar P (t_1) + \sum_{k=2}^n \bar P ( t_{k} ) P (t_{k-1}) \cdots P(t_2) P(t_1)
\nonumber \\
&+& P (t_n) \cdots P(t_2) P(t_1)
\label{23}
\eea
We thus identify the first crossing class operator as
\beq
C_k =  \bar P ( t_{k} ) P (t_{k-1}) \cdots P(t_2) P(t_1)
\label{24}
\eeq
for $ k \ge 2 $, with $C_1 = \bar P (t_1)$. This clearly describes histories
which are in $ x>0$ at times $ t_1, t_2, \cdots t_{k-1}$ and in $x<0$ at
time $t_k$, so, to within the limits of the Zeno effect outlined above,
describe a first crossing between $t_{k-1}$ and $t_k$. The last term
in Eq.(\ref{23}) is the non-crossing class operator, $C_{nc}$. We will
actually assume that $\tau$ is sufficiently large that the wave packet
ends up entirely in $x<0$ at large times, to $C_{nc} | \psi \rangle $
is essentially zero. This means that we effectively have
\beq
\sum_{k=1}^n C_k = 1
\eeq
as required. We will generally be interested in class operators describing
crossing in intervals $[t_{\a}, t_{\a+1}] $ of size $ \Delta = m \epsilon$,
where $ m $ is a positive integer, and these class operators
are simply obtained by summing,
\beq
C_{\a} = \sum_{k \in \alpha} C_k
\eeq
Last crossing class operators are similarly constructed but will not be required,
as we shall see below.

The class operators Eq.(\ref{24}) are difficult to work with analytically,
so some sort of simplification or approximation is necessary. One possible approach is
to make use of the result of Echanobe et al. \cite{Ech}, who showed that the unitary
evolution interspersed with projections in Eq.(\ref{17}) is approximately the
same (up to overall unitary factors) as evolution with a Hamiltonian including a complex potential of the
form $V (x) = - i V_0 \theta ( -  x ) $. This very interesting possibility
is explored in detail in \cite{HaYe1}. Here, we will work directly with Eq.(\ref{24})
and use a simple semiclassical approximation.

Consider the strings of identical projection operators $P$ at different
times appearing in Eq.(\ref{17}) (and similarly in Eq.(\ref{24})).
Given that the final projection $P(t_n)$ is onto $x>0$ and also that the initial
state is localized in $x>0$, it seems reasonable to suppose that
the projections at times $t_1$ to $t_{n-1}$ do not disturb the
evolving state too much, under the condition $ \e > t_z $ discussed above.
It therefore seems
reasonable to make the approximation
\beq
 P (t_n) \cdots P(t_2) P(t_1) | \psi \rangle \approx P (t_n) | \psi \rangle
\label{21}
\eeq
This is easy to understand in a path integral representation. The right-hand side
is in essence the amplitude from an initial state concentrated in $x>0$ to a final point
in $x>0$ at time $t_n$ (up to overall unitary factors). The sum over paths will be dominated by the straight line
path, which lies entirely in $x>0$ at all intermediate times. It will therefore
be little affected by the insertion of additional projections onto $x>0$
at intermediate time.

Using this approximation, the crossing class operator Eq.(\ref{24}) operating
on the given initial state may
be approximated as
\beq
C_{k} \approx  \bar P ( t_{k+1} ) P (t_{k})
\eeq
Rearranging and using the approximation Eq.(\ref{21}) a second time we obtain
\bea
C_{k} &=& P ( t_k ) - P ( t_{k+1})  P ( t_k )
\nonumber \\
& \approx & P ( t_k ) - P ( t_{k+1} )
\eea
By summing over an appropriate range of $k$, it is easily seen that the coarser-grained
class operator $C_{\a}$ for crossing during a time interval $[t_{\a}, t_{\a+1}]$
of size $\Delta $ is
\bea
C_{\a} &=& P( t_{\a} ) - P( t_{\a+1} )
\nonumber \\
&=& \int_{t_\a}^{t_{\a+1}} dt \ \frac { (-1) } {2 m } \left( \hat p \delta ( \hat x ) + \delta (\hat x ) \hat p \right)
\label{30}
\eea
This is precisely of the anticipated form, Eq.(\ref{2}). Hence the expected class operator
arises naturally in a simple semiclassical approximation. Gratifyingly, despite the rather
crude nature of this semiclassical approximation, this result coincides with the complex
potential calculation of Ref.\cite{HaYe1}, as long as the timescale $\Delta$ is sufficiently
large, compared to the natural timescale of the complex potential, $1/ V_0$.
We also note that the semiclassical approximation used above means that there is no
distinction between first and last crossing, so a last crossing class operator
would yield the same result.

It now remains to check for decoherence of histories for some interesting initial
states.
We consider the particular case of an initial state consisting of a wave packet
\beq
\psi (x) = \frac{1} {(2 \pi \sigma^2)^{1/4} } \exp \left( - \frac {(x-q_0)^2} {4 \sigma^2 } + i p_0 x \right)
\label{7.33}
\eeq
where $q_0 > 0 $ and $ p_0 < 0 $.
In the simplest case, the wave packet crosses
the origin almost entirely during one of the time interval $ [t_\a, t_{\a+1}]$ for some
fixed $\a$, without
any substantial overlap with any other time intervals. This means
that
\bea
C_{\a} | \psi \rangle &  \approx & | \psi \rangle
\nonumber \\
C_{\b} | \psi \rangle  & \approx & 0 \ \ {\rm for } \ \ \b \ne \a
\eea
and it follows that $D (\a, \b) \approx 0 $. The key time scales here
are the classical arrival time for the centre of the packet,
\beq
t_a = \frac { m q_0 } { | p_0 | }
\eeq
and the Zeno time Eq.(\ref{18}), which for the wave packet Eq.(\ref{7.33}) is
\beq
t_z = \frac { m \sigma} { | p_0 | }
\eeq
(up to irrelevant constants).
Here, the Zeno time
is seen to be the time taken for the wave packet to move a distance equal to it spatial
width $\sigma$, or equivalently, it is the size of the packet's ``temporal
imprint'' at the origin. Therefore, the above approximations work if,
firstly,
\beq
\Delta  \ \gg \ t_z
\eeq
and secondly, if the classical arrival time $t_a$ lies inside
the interval $[t_\a, t_{\a+1}]$ and is at least one or two Zeno times
away from the boundaries.

For smaller values of $\Delta$, a given packet will encounter
the origin in a number of different time intervals, so more than
one of the crossing states $C_{\a} | \psi \rangle $ will be non-zero,
leaving the possibility that some of them may be non-orthogonal.
The underlying physical effect is ``diffraction in time''
\cite{diff} and will be explored elsewhere in more detail. A detailed calculation
in Ref.\cite{HaYe1} indicates that with the above initial state, there is still
good approximate decoherence
of histories for values of $\Delta $ of order $t_z $ or less, as long as
the wave packet is sufficiently strongly peaked in momentum,
$ | p_0 | \sigma \gg 1 $. (Note that such small $\Delta$ does not fall
foul of the Zeno effect discussed above since the possibility of
reflection is effectively ignored once we are in the regime of the
semiclassical approximation Eq.(\ref{21})).

It is easy to see that
similar conclusions hold for superpositions of initial states of the
form Eq.(\ref{7.33}) as long as they are approximately orthogonal.
Loosely, this is because under the above conditions,
the class operators do not disturb the states and the only non-zero
components of the off-diagonal terms of the decoherence functional
will be proportional to the overlap of pairs of initial wave packets,
so will be approximately zero. More general, non-orthogonal superpositions
may, however, produce backflow, so there may be no decoherence.

In summary, the decoherent histories approach to the arrival time problem exposes
the standard result Eq.(\ref{1}) as a quasi-probability, valid when there is decoherence
of histories, but not otherwise. In particular, its negativity when there is backflow
corresponds precisely to non-decoherent situations.

\section*{Acknowledgements}

We are grateful to Gonzalo Muga and Larry Schulman for useful
discussions. JJH acknowledges the hospitality of
the Max Planck Institute in Dresden at which some of this work was
carried out during the Advanced Study Group, ``Time: Quantum and Statistical Mechanics Aspects",
August 2008.

\bibliography{apssamp}

\end{document}